\documentclass{article}
\usepackage{spconf,amsmath,graphicx,xcolor,balance,url}

\title{JOINT BLIND ROOM ACOUSTIC CHARACTERIZATION FROM SPEECH AND MUSIC SIGNALS USING CONVOLUTIONAL RECURRENT NEURAL NETWORKS}
\name{Paul Callens$^{1,2}$, Milos Cernak$^2$}
\address{
  $^1$\'{E}cole Polytechnique F\'{e}d\'{e}rale de Lausanne (EPFL), 1015, Lausanne, Switzerland \\
  $^2$Logitech Europe S.A., 1015, Lausanne, Switzerland
}
\begin{document}
%
\maketitle
\begin{abstract}
Acoustic environment characterization opens doors for sound reproduction innovations, smart EQing, speech enhancement, hearing aids, and forensics. Reverberation time, clarity, and direct-to-reverberant ratio are acoustic parameters that have been defined to describe reverberant environments. They are closely related to speech intelligibility and sound quality. As explained in the ISO3382 standard, they can be derived from a room measurement called the Room Impulse Response (RIR). However, measuring RIRs requires specific equipment and intrusive sound to be played. The recent audio combined with machine learning suggests that one could estimate those parameters blindly using speech or music signals. We follow these advances and propose a robust end-to-end method to achieve blind joint acoustic parameter estimation using speech and/or music signals. Our results indicate that convolutional recurrent neural networks perform best for this task, and including music in training also helps improve inference from speech.

\end{abstract}

\section{INTRODUCTION}
\label{sec:intro}

A source sound signal $x(t)$ coming from a speaker is subject to reverberations and noise $n(t)$ when played in a room. The resulting signal can be expressed as:
$$y(t)=x(t)*h(t) + n(t)$$
with $h(t)$ the room impulse response (RIR). RIR can be measured with the sine-sweep technique \cite{Farina2000SimultaneousTechnique}. It consists of emitting a logarithmic sine-sweep sound with a speaker to excite every frequency one after another. A microphone picks up the direct and reverberant signals from another spot in the room from which the room impulse response is reconstructed.

Standard acoustic parameters \cite{BSENISO3382-1:20092009AcousticsSpacesb} have been derived from the RIR to describe the characteristics of a room with measures closer to human perceptions. We focused on room characteristics that describe how pleasant music will sound or speech will be intelligible:
(i) \textit{Reverberation time (RT60)} is defined by the time it takes for the sound energy to decay 60dB after the source is switched off; (ii) \textit{Clarity (C50/C80)} is measured by calculating the ratio between the early reflections energy (up to 50/80ms) and the energy of the late response from the decay curve; (iii) \textit{ Direct-to-Reverberant Ratio (DRR)}, which, similarly to C50, is a ratio of the direct sound energy over the later energy, considering that the direct sound is included in the leading 2.5  milliseconds of the RIR.

Many attempts to characterize room acoustics were conducted in the last decade. In 2015, the ACE challenge~\cite{eaton2016estimation} was organized to stimulate research and establish
state of the art in the domain. The majority of the proposed techniques were non-machine-learning based methods. In 2018, Gamper and Tashev used Convolutional Neural Networks (CNNs) and Gammatone filterbanks to predict the average RT60 of a reverberant signal~\cite{gamper2018blind}, and outperformed the best ACE challenge method. Recent work of Looney and Gaubitch~\cite{looney2020joint} showed promising results in the joint blind estimation of RT60, DRR, and Signal-to-Noise Ratio (SNR). The majority of previous studies used the input speech signals only. Music was considered only in a few other cases~\cite{kendrick2008monaural,peters2012name}.

Instead of room acoustic parameter prediction, some works try to classify the rooms directly. For example, recent comparison~\cite{papayiannis2018end} of different deep neural network architectures: CNNs, Recurrent Neural networks, and Convolutional-Recurrent
Neural Networks (CRNNs) on the classification of 7 separate rooms from the ACE challenge
using reverberant speech showed that the CRNN architecture performed the best.

All previous studies of blind acoustic environment characterization explored mostly CNNs architectures. In this work, 
we try to find the decay of a signal, and it makes sense considering the sequence aspect of data, and thus using recurrent layers might be beneficial. Besides, we aim to perform a joint estimation of RT60, C50, C80, and DRR parameters and using individual or mixed input speech and music signals. To our knowledge, using deep learning for acoustic room characterization from music has not been studied.

The paper is structured as follows: Section~\ref{sec:metho}
introduces the proposed methods, Section~\ref{sec:framework} describes experimental setup, Section~\ref{sec:results} presents the results and Section~\ref{sec:conclusion} discusses the results and concludes the paper.

\begin{figure*}[htbp]
    \centering
    \includegraphics[width=0.9\linewidth]{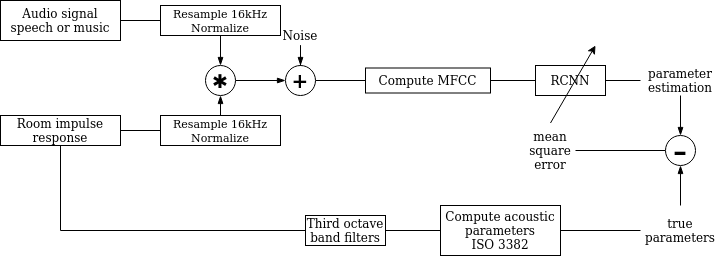}
    \caption{Overall acoustic parameter estimation pipeline.}
    \label{fig:pipeline}
\end{figure*}

\section{Proposed methods}
\label{sec:metho}

CRNN architectures have been used to estimate room acoustic parameters from pre-processed reverberant and noisy audio inputs. The research method consisted of 3 main steps: i) data preparation, ii) data pipeline incl. input and output feature extraction and normalization, and iii) designing and training neural networks.

\subsection{Data preparation}

The methods used to compute acoustic parameters from room impulse responses have some limitations. In our analysis, some RT60 values were too high to be real (over 10 seconds). These errors are probably due to too high background
noise. We thus discarded RIRs with reverberation times above four seconds. In reality, four seconds is about the reverberation of a church, which can be considered extreme.

All the audio samples are normalized, each using min-max normalization, assuming that the min of an audio signal corresponds to silence and thus should be zero: $S_{norm} = S / max(|S|)$.

The duration of input audio samples was a hard choice to make and required many trials. We tried to find a good compromise between computation time (that would get larger for longer signals), model complexity, and evaluation score. After many tests, we decided to use the 8 seconds chunks of the music and speech signals from our dataset,
as they were a good compromise for both music and speech. The beginning of the RIR recordings was trimmed until the onset for more
uniformity within the dataset.

\subsection{Data pipeline}

Figure~\ref{fig:pipeline} shows the pre-processing pipeline. The true acoustic values are extracted from the RIR recordings to give us true parameter values for the network output. Besides, the audio signals are processed to simulate reverberation and noise. Lastly, a spectral representation of the signal is computed to be used as input for the neural networks.

The true acoustic parameters are computed as defined in the International Standard of Room Acoustic Measurements, ISO3382~\cite{international2009acoustics}. The ISO standard suggests to perform a linear least-square fit to the -5dB to -35dB. That yields RT30 (the time it takes for SPL to drop 30dB). It is then extrapolated to RT60 by multiplying by a factor of 2. For each parameter, frequency-dependent values were computed and then averaged. To get frequency-dependent acoustic parameters, we applied a band-pass filter to the RIR signals before deriving RT60, C50, C80, and DRR. The center frequency bands for those filters were chosen by octaves accordingly to ISO 266:1997 standard~\cite{iso1997266} between 125Hz and 4kHz. The higher octave, 8kHz, would not allow applying
a band-pass filter to satisfy the Nyquist sampling criterion with 16kHz samples. Finally, acoustic parameters were found for the following octave bands: 125, 250, 500, 1K, 2K, and 4K.

\begin{figure}[htbp]
    \centering
    \includegraphics[width=0.7\linewidth]{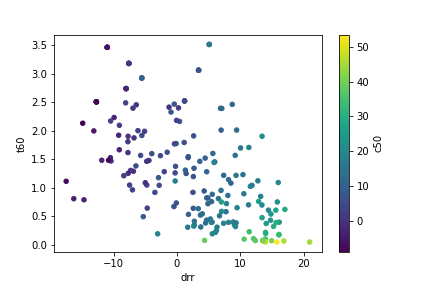}
    \caption{Distribution of the measured acoustic parameters}
    \label{fig:ac_distrib}
\end{figure}

Figure \ref{fig:ac_distrib} shows the distribution of the measured acoustic parameters from our set of RIRs after it has been balanced. It also highlights the correlation between T60, DRR, and C50, which can be interpreted as the fact that a room with a high reverberation goes along with lower clarity and direct-to-reverberant ratio; thus, the speech will be harder to understand.

\subsection{Evaluation metrics}


Two metrics are computed to compare the models. The Mean Absolute Percentage Error (MAPE) and the Root Mean Square Error (RMSE):

$$\epsilon_M = \text{MAPE} = \frac{100\%}{n}\sum_{i=1}^n \left\lvert\frac{y_{pred}-y_{true}}{y_{true}}\right\rvert[\%]$$

$$\epsilon_R = \text{RMSE} = \sqrt{\sum_{i=1}^n \frac{(y_{pred}-y_{true})^2}{N}}[s\text{ or } dB]$$



\section{Experimental framework}
\label{sec:framework}

\subsection{Data}
\label{sec:data}

The dataset is composed of music and speech, collected from the MUSAN dataset \cite{Snyder2015MUSAN:Corpus}, and RIRs, gathered from online open-source libraries : OpenAirLib \cite{Brown2017OpenAirLib:Spaces}, echo-thief \cite{WarrenEchoThiefUrl}, MIT IR Survey \cite{Traer2016MITDataset}, and RWCP \cite{Nakamura2000AcousticalRecognition}. Fifteen RIR measurements have also been made in-house for testing purposes. From these same RIRs, acoustic parameters are computed to be used as reference values for algorithm training. The MUSAN data is split into 80/20\% training and testing subsets, respectively. The 406 collected RIRs were split into 306 used for training and 100 for testing.

All the data is normalized and downsampled to mono-channel 16kHz. The speech and music audio files are trimmed to 8 seconds. Reverberation is simulated by convolving speech and music with RIRs and adding noise. Random Gaussian noise is added to the reverberant signal at the SNRs of 15, 10, 5, and 0 dB. Mel Frequency Cepstral Coefficients (MFCC) features are computed from noisy reverberant audio and used as input for the neural network. For our experiments, we used 25ms frame size and 10ms frame steps. The MFCCs were obtained using Librosa~\cite{mcfee2015librosa}.

\subsection{Neural Network}
\label{subsec:nn}

Table~\ref{tab:conv-params} shows the architectures of the baseline and two proposed models. The baseline~\cite{looney2020joint} represents the state-of-the-art model for the joint blind estimation of acoustic parameters, and the CRNN1 adds two more recurrent layers. Both the baseline and CRNN1 use the ReLU activation functions. The number of trainable parameters is rather high, and thus we propose the CRNN2 model that has almost five times fewer parameters, and is more suitable for embedded platforms.

\begin{table}[htbp]
\centering
\caption{Model architectures of the baseline~\cite{looney2020joint} and two our proposed CRNNs. The output classes are RT60, C50, C80, and DRR. ``C`` stands for Conv2D, ``G`` for GRU, and ``D`` for Dense layers.}\vspace{9pt}
\label{tab:conv-params}
\begin{tabular}{c|c|c|c}
 & \multicolumn{3}{c}{(size) / (kernel size, number of filters)} \\ \hline
Model & Baseline & CRNN1 & CRNN2 \\
layers & \#1.66M & \#1.74M & \#369K \\ \hline 
1 & C(5, 256) & C(5, 256) & C(3, 64) \\ \hline
2 & C(5, 256) & C(5, 256) & C(3, 128) \\ \hline
3 & D(64) & G(64) & C(3, 128) \\ \hline
4 & D(4) & G(64) & C(3, 128) \\ \hline
5 & - & D(64) & G(32) \\ \hline
6 & - & D(4) & G(32) \\ \hline
7 & - & - & D(128) \\ \hline
8 & - & - & D(64) \\ \hline
9 & - & - & D(4) \\
\end{tabular}
\end{table}

Batch normalization, dropout, and max pooling methods is used with the Conv2D layers. The CRNN2 model uses the ELU activation function. To optimize the training process of the all three models, batch training with the size of 64, early stopping with patience set to 15, ADAM optimizer~\cite{kingma2014adam} with the learning rate $\alpha=0.001$, and the mean square error loss is implemented.

\section{Results}
\label{sec:results}

\begin{figure*}[htbp]
    \centering
    \includegraphics[width=1.0\linewidth]{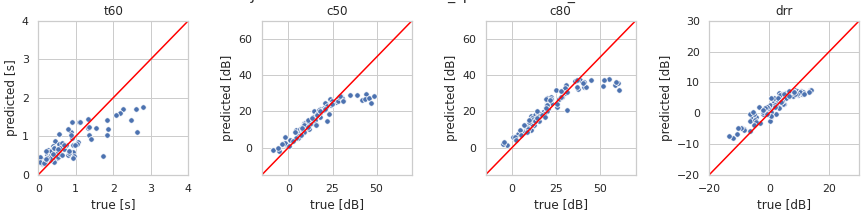}
    \caption{Performance of the joint blind estimation on audio, mixed speech and music, data.}
    \label{fig:scatteraudio}
\end{figure*}

Table~\ref{tab:jspeech} shows the evaluation of the joint blind estimators on the training and testing speech signals. Both proposed models outperform the baseline for the T60, C50, and C80 parameters. Except for the DRR parameter, the CRNN2 performs in most cases the best.

\begin{table}[h]
\caption{Speech estimation errors $\epsilon_M$ and $\epsilon_R$ obtained with the speech estimator.}\vspace{9pt}
\label{tab:jspeech}
\centering
\begin{tabular}{l|c|c|c|c|c|c|c|c}
    & \multicolumn{2}{c|}{T60} & \multicolumn{2}{c|}{C50} & \multicolumn{2}{c|}{C80} & \multicolumn{2}{c}{DRR}\\ \cline{2-9}
    Method & $\epsilon_M$ & $\epsilon_R$ & $\epsilon_M$ & $\epsilon_R$ & $\epsilon_M$ & $\epsilon_R$ & $\epsilon_M$ & $\epsilon_R$\\
    \hline
    Baseline & 65 & 0.4 & \textbf{41} & 6.7 & 34 & 8.5 & \textbf{116} & \textbf{3.0}\\
    CRNN1 & 55 & 0.4 & 42 & \textbf{6.5} & 33 & \textbf{8.2} & 126 & 3.2\\
    \textbf{CRNN2} & \textbf{54} & 0.4 & \textbf{41} & \textbf{6.5} & \textbf{32} & \textbf{8.2} & 120 & 3.1\\
\end{tabular}
\end{table}

Table~\ref{tab:jmusic} shows the evaluation of the joint blind estimators on the training and testing music signals. The CRNN1 model outperforms the baseline. In contrast to the speech evaluation, the CRNN2 performs competitively for the music signals only to estimate clarity parameters.

Figure~\ref{fig:scatteraudio} shows the performance of the joint blind estimation on audio (mixed speech and music) data. We see that the estimates degrade with more challenging environments with stronger late reverberation. Table~\ref{tab:jaudiospeech} shows the evaluation of the joint audio estimators on the speech signals. Surprisingly, we see a significant positive effect of training music and speech together; this system even outperforms the estimator train on pure speech (Table~\ref{tab:jspeech}).

\begin{table}[h]
\caption{Music estimation errors $\epsilon_M$ and $\epsilon_R$ obtained with the music estimator.}\vspace{9pt}
\label{tab:jmusic}
\centering
\begin{tabular}{l|c|c|c|c|c|c|c|c}
    & \multicolumn{2}{c|}{T60} & \multicolumn{2}{c|}{C50} & \multicolumn{2}{c|}{C80} & \multicolumn{2}{c}{DRR}\\ \cline{2-9}
    Method & $\epsilon_M$ & $\epsilon_R$ & $\epsilon_M$ & $\epsilon_R$ & $\epsilon_M$ & $\epsilon_R$ & $\epsilon_M$ & $\epsilon_R$\\
    \hline
    Baseline & 139 & 0.6 & 93 & 11.9 & 85 & 14.5 & 197 & 5.2\\
    \textbf{CRNN1} & \textbf{126} & 0.6 & 87 & \textbf{10.9} & 83 & \textbf{13.3} & \textbf{193} & \textbf{5.0}\\
    CRNN2 & 152 & 0.6 & \textbf{80} & 12.0 & \textbf{72} & 14.6 & 208 & 5.7\\
\end{tabular}
\end{table}


\begin{table}[h]
\caption{Speech estimation errors $\epsilon_M$ and $\epsilon_R$ obtained with the audio (trained on mixed speech + music) estimator.}\vspace{9pt}
\label{tab:jaudiospeech}
\centering
\begin{tabular}{l|c|c|c|c|c|c|c|c}
    & \multicolumn{2}{c|}{T60} & \multicolumn{2}{c|}{C50} & \multicolumn{2}{c|}{C80} & \multicolumn{2}{c}{DRR}\\ \cline{2-9}
    Method & $\epsilon_M$ & $\epsilon_R$ & $\epsilon_M$ & $\epsilon_R$ & $\epsilon_M$ & $\epsilon_R$ & $\epsilon_M$ & $\epsilon_R$\\
    \hline
    Baseline & 63 & 0.4 & 48 & 6.9 & 45 & 8.9 & 132 & 3.3\\
    CRNN1 & 55 & 0.4 & 43 & \textbf{6.4} & 37 & 8.1 & 138 & 3.3\\
    \textbf{CRNN2} & \textbf{50} & \textbf{0.3} & \textbf{42} & \textbf{6.4} & \textbf{34} & \textbf{8.0} & \textbf{127} & \textbf{3.2}\\
\end{tabular}
\end{table}


\subsection{Robustness to noise}

All results reported in the previous section were averaged for all considered SNRs. However, the performance was stable for all noise conditions. For example, the T60 $\epsilon_M$ was 50, 49, 49, and 51, for the SNRs of [15, 10, 5, 0] dB, respectively.

\section{Discussion and conclusion}
\label{sec:conclusion}

We proposed an end-to-end method to blindly and jointly estimate four acoustic parameters, namely reverberation time (RT60), direct-to-reverberant ratio (DRR), and clarity (C50 and C80). Two kinds of input signals were investigated: speech and music. Our explorations focused on two novel aspects of the joint estimation of acoustic parameters for room characterization. First, novel CRNN architectures have been designed that outperform current state-of-the-art systems.
Our results showed that using recurrent layers is beneficial for the explored task.
Second, models were trained to estimate parameters using music signals, which, contrary to speech signals, have not been subject to many studies. Finally, we showed that combining music with the speech signals in training improves the speech signals' inference.

However, the results show that music is less suitable for parameter estimation. One possible explanation is that the notes are tempered on a specific scale. Contrary to speech, music played by instruments usually does not stop sharply. Thus, fewer decay curves are available, which makes the task harder. Moreover, reverberation is a widely used effect in music production. This effect, used with moderation, generally has a positive impact on our music experience. It is likely that this effect added up with the convolution we applied to music signals, which introduced bias in the estimations.

The music, speech, and RIR samples were carefully split for training and testing, and a mixture of datasets has been used. This care was taken to ensure the validity of the results and our algorithm's robustness against noise and unseen examples. However, the validity of impulse response measurements can be discussed. The analysis of available datasets containing multiple acoustic measurements for one room showed that there could be considerable variations between acoustic parameter values within the same room. Thus, our true values for acoustic parameters are also subject to uncertainty, reflecting on the estimations.

For this research, more than 160 models were trained, with more than 1TB of audio data. During these explorations, some parameters have been chosen by intuition and flair. In the future, we would like to focus on the estimation of other characteristics of a room like its volume or the speaker's distance to the closest wall. 

\section{Acknowledgements}
This work was realized during a six-month internship at Logitech.

\vfill\pagebreak

\balance
\bibliographystyle{IEEEbib}
\bibliography{strings,refs,references}

\end{document}